\documentclass{jpp}
\usepackage{graphicx,natbib}
\usepackage{epstopdf, epsfig}
\usepackage{amsfonts,amsmath,amssymb}
\usepackage[pdftex,colorlinks,citecolor=blue]{hyperref}
\usepackage{bm}
\usepackage{xcolor}
\usepackage[]{cleveref}%capitalise, noabbrev
\crefname{section}{\S\!}{\S\S\!}
\Crefname{section}{Section}{Sections}
\crefname{appendix}{App.}{Apps.}
\Crefname{appendix}{Appendix}{Appendices}
\crefname{equation}{Eq.}{Eqs.}
\Crefname{equation}{Equation}{Equations}
\crefname{figure}{Fig.}{Figs.}
\Crefname{figure}{Figure}{Figures}
\newcommand{\mockalph}[1]{}	

\newcommand{\orcid}[1]{}

\newcommand{\bh}{\hat{\bm{b}}}
\newcommand{\B}{\bm{B}}
\newcommand{\dB}{\delta \bm{B}}
\newcommand{\mB}{\overline{\bm{B}}}
\newcommand{\mBm}{B_{0}}
\newcommand{\va}{\bm{v}_{\rm A}}

\newcommand{\vam}{{v}_{\rm A}}
\newcommand{\ph}{\hat{\bm{p}}}
\newcommand{\alfreds}{M+21}

\newcommand{\revchng}[1]{{#1}}

\ifCUPmtlplainloaded \else
  \checkfont{eurm10}
  \iffontfound
    \IfFileExists{upmath.sty}
      {\typeout{^^JFound AMS Euler Roman fonts on the system,
                   using the 'upmath' package.^^J}%
       \usepackage{upmath}}
      {\typeout{^^JFound AMS Euler Roman fonts on the system, but you
                   dont seem to have the}%
       \typeout{'upmath' package installed. JPP.cls can take advantage
                 of these fonts, if you use 'upmath' package.^^J}%
       \providecommand\upi{\upi}%
      }
  \else
    \providecommand\upi{\upi}%
  \fi
\fi

\ifCUPmtlplainloaded \else
  \checkfont{msam10}
  \iffontfound
    \IfFileExists{amssymb.sty}
      {\typeout{^^JFound AMS Symbol fonts on the system, using the
                'amssymb' package.^^J}%
       \usepackage{amssymb}%
         \let\leq=\leqslant
         
      }{}
  \fi
\fi

\ifCUPmtlplainloaded \else
  \IfFileExists{amsbsy.sty}
    {\typeout{^^JFound the 'amsbsy' package on the system, using it.^^J}%
     \usepackage{amsbsy}}
    {\providecommand\boldsymbol[1]{\mbox{\boldmath $##1$}}}
\fi

\title[]{On the construction of general large-amplitude spherically polarised Alfv\'en waves  }

\author[J.~Squire and others]%
{J.~Squire\orcid{0000-0001-8479-962X}$^{1}$\thanks{Email address for correspondence: jonathan.squire@otago.ac.nz},
 A.~Mallet\orcid{0000-0001-9202-1340}$^{2}$}

\affiliation{$^1$Physics Department, University of Otago, Dunedin 9010, New Zealand\\
$^2$Space Sciences Laboratory, University of California, Berkeley, CA 94720, USA}

\pubyear{2022}
\volume{}
\pagerange{}
\date{}

\begin{document}
\maketitle

\begin{abstract}

In a magnetised plasma on scales well above ion kinetic scales,  any constant-magnitude magnetic field, accompanied
by parallel Alfv\'enic flows, forms a nonlinear solution in an isobaric, constant-density background. These structures, which are also known as spherically polarised Alfv\'en waves, are observed ubiquitously in the solar wind, presumably created by the growth of small-amplitude fluctuations as they propagate outwards in the corona. Here, we present a computational method to construct such solutions 
of arbitrary amplitude with general multi-dimensional structure,
and explore some of their properties. The difficulty lies in computing a zero-divergence, constant-magnitude  magnetic field, 
which leaves a single, quasi-free function to define the solution, while requiring
strong constraints on any individual component of the field. Motivated by the physical process of wave growth in 
the solar wind, our method  circumvents this issue by starting from low-amplitude Alfv\'enic fluctuations dominated by a strong mean field, then ``growing''
 magnetic perturbations  into the large-amplitude regime. We present example solutions with nontrivial structure in one, two, and three dimensions, demonstrating 
a clear tendency to form very sharp gradients or discontinuities, unless the solution is one dimensional. 
As well as being useful as an input for other calculations, particularly  the study of parametric decay, 
our results provide a natural explanation for the extremely sharp field discontinuities observed across magnetic-field switchbacks
in the low solar wind.

\end{abstract}

\section{Introduction}

The existence of the incompressible Alfv\'en wave is perhaps the most important distinguishing feature 
between the dynamics of neutral fluids and plasmas, underlying key physics of turbulent energy dissipation
and enabling nontrivial dynamics well below the scale of the mean-free path \citep[e.g.,][]{Schekochihin2009}. 
Interestingly---and unlike other wave-like perturbations of fluids or plasmas---Alfv\'en waves have an unambiguous
nonlinear generalisation \citep{Barnes1974}: defining the plasma density $\rho$, pressure $P$, magnetic-field $\bm{B}$, flow velocity 
$\bm{u}$, and field strength $B=|\bm{B}|$, any perturbation that satisfies
\begin{equation}
    P  = {\rm const.},\quad\rho= {\rm const.},\quad B^2 = {\rm const.},\quad   \delta\bm{u} = \pm \delta \bm{B}/\sqrt{4\pi \rho} \label{eq: nl aw solution}
\end{equation} 
is a nonlinear solution that propagates in the direction $\bh = \bm{B}/B$ at speed $\vam \equiv |\mB|/\sqrt{4\pi\rho}$ (where 
the overline signifies a spatial average, and $\delta \bm{B} =\B-\mB$). \revchng{Such solutions are necessarily maximally ``imbalanced'' (the magnitude of the fluctuation's cross helicity equals their energy), and, in the wave frame at speed $\vam$, involve constant plasma kinetic energy (constant $|\bm{u}|$) and zero motional electric field $\bm{u}\times \bm{B}$ \citep{Matteini2015}. } Moreover, unlike large-amplitude sound or magnetosonic waves, spherically polarised 
perturbations do not steepen into shocks even if $|\delta \bm{B}|\gg |\mB|$, although they can be unstable \citep{Sagdeev1969,Cohen1974a}.
They also share many properties of the linear (small-amplitude) Alfv\'enic fluctuations, such as how
they change in amplitude and/or refract in a nonhomogenous background medium \citep[e.g.,][]{Barnes1974,Hollweg1974}. 
The solution \eqref{eq: nl aw solution} is even valid in a collisionless plasma as well as in collisional magnetohydrodynamics (MHD), so long as $\delta \bm{B}$  varies over scales much larger than the ion gyroradius or skin depth \citep{Barnes1971,Kulsrud1983}. %\footnote{The constant-$P$ condition must be replaced by separate conditions on the parallel and perpendicular pressures, with the relation between $\bm{u}$ and $\delta \B$ appropriately modified to account for the mean pressure anisotropy.} 
Given these properties, it is perhaps not surprising that  perturbations close to \cref{eq: nl aw solution} are observed ubiquitously in 
our best-studied example of a natural plasma---the solar wind  \citep{Belcher1971}. In this context, 
they are  known as \emph{spherically polarised} and  their properties likely underly a number of key aspects of solar-wind physics, including
 heating and acceleration \citep[e.g.,][]{Shoda2021,Bale2021}, turbulent spectra \cite[e.g.,][]{Matteini2018,Bowen2021}, and properties of ``switchback'' field reversals \citep{Kasper2019,Squire2020,Johnston2022}.

Despite these important features, to our knowledge, there does not exist any general method 
to construct and study such nonlinear Alfv\'enic solutions when $\delta \B\gtrsim |\bm{B}|$, or 
even any general results regarding their existence. Previous methods  (\citealp{Valentini2019}; \citealp[see also][]{Roberts2012}), though promising, may lead to unnecessary discontinuities in the solutions, the causes of which are discussed below. The difficulty arises from the twin constraints 
of $\nabla\cdot\B = 0$ and $B=|\bm{B}|={\rm const.}$, which leaves only one degree of freedom from which to 
construct the 3-D vector field $\bm{B}$. Moreover, as we show below, this degree of freedom cannot in general 
be freely chosen when $\delta \B$  approaches $\mB$ in magnitude, even though simple, smooth 
solutions do exist (they just require nontrivial constraints on all components of $\bm{B}$ simultaneously).  
It is the purpose of this letter to present such a method and briefly explore the properties of 
the solutions \eqref{eq: nl aw solution}.
These are of interest for two reasons. First, the method can be used as  input or initial conditions for other numerical calculations, such as the study of parametric decay \citep{DelZanna2001,Primavera2019} or large-amplitude reflection-driven turbulence \citep{Squire2020,Johnston2022}. 
Second, the solutions are interesting in and of themselves---our method
shares strong similarities with the plasma-expansion process that generates large-amplitude Alfv\'enic states
in the solar wind, so the structures and properties that arise may be of direct physical relevance. 
Of particular interest is the development of sharp field discontinuities, which we show form much more readily 2-D or 3-D
solutions than in 1-D; this is promising, since highly discontinuous fields seem to be observed in switchbacks by Parker Solar Probe and other spacecraft \citep{Bale2019,AkhavanTafti2021}.
More generally, the method provides an extremely broad class of nonlinear solutions 
to the MHD equations that complements other  solutions such as force-free magnetostatic equilibria  \citep{Marsh1996}. Seen differently, it provides  a way to construct a general divergence-free, unit-vector field. 

Below, we first discuss the basic idea of the method and equations involved, then present its 
application to magnetic-field configurations that vary in one, two, and three dimensions, respectively. 
The one-dimensional version, although idealised and explored previously  in \citet[][hereafter \alfreds]{Mallet2021} and \citet{Squire2022}, provides 
a helpful illustration of the method and the challenges involved in higher dimensions. We finish with brief discussion
of some interesting features of the solutions, focused particularly on the generation of sharp gradient and/or discontinuities, as well as some possible applications of our results. 

\section{Method}

The method we propose is based on splitting the magnetic field into its mean ($\mB$) and fluctuating ($\dB$) parts, 
then devising an equation to grow $\dB$ in amplitude while maintaining $\overline{\dB}=0$, $\nabla\cdot\dB=0$,  fixed $\mB$, and constant $B^{2} = |\mB + \dB|^{2}$.
In this way, the method ``grows'' a small amplitude spherically polarised wave, which can be relatively easily constructed from linear Alfv\'enic perturbations via standard optimization
methods, to any desired amplitude $\mathcal{A}\equiv (\overline{\dB^{2}}/\mB^{2})^{1/2}$. 
The obvious candidate to evolve $\dB$ is simply the induction equation supplemented by exponential growth,
\begin{equation}
\frac{\partial}{\partial t} \dB = \dB + \nabla\times [\tilde{\bm{u}}\times (\mB+\dB)],\label{eq: induction}
\end{equation}
which clearly maintains $\overline{\dB}=0$ and $\nabla\cdot\dB=0$ by construction. The flow $\tilde{\bm{u}}$ should be non-Alfv\'enic: this 
is not the flow associated with the Alfv\'en wave itself ($\delta \bm{u}$ in \cref{eq: nl aw solution}), but rather the flow needed to change the
shape of $\dB$ as it grows with $t$, in order to maintain constant $B$. The clear choice is the potential/compressive flow $\tilde{\bm{u}} = \nabla\phi$, 
which  cannot contribute to the Alfv\'enic flow on a periodic domain because $\nabla\cdot\tilde{\bm{u}}=\nabla^{2}\phi$.
To proceed, we form ${\partial}/{\partial t}|\mB + \dB|^{2} = 2(\mB + \dB)\cdot {\partial \dB}/{\partial t}$ and assert that its fluctuating part 
must be zero, \emph{viz.,} that $\phi$ should be chosen so that $\partial/\partial t\,\delta ( B^{2})= 0$.  
Using the fact that $\nabla B^{2}=0$ and $\nabla \mB = 0$, this yields
\begin{equation}
B^{2} \nabla_{\perp}^{2}\phi = B^{2}\nabla^{2}\phi - \sum_{ij}B_{i}B_{j}\nabla_{i}\nabla_{j}\phi = -\dB\cdot\mB.\label{eq: phi constraint},
\end{equation}
where $B_{i} = \overline{B}_{i} + \delta B_{i}$.
If \cref{eq: phi constraint} can be solved at each step, with the result used to evolve $\dB$ through \cref{eq: induction}, the
total field will maintain spatially constant $B^{2}$ (and $\nabla\cdot\bm{B} = 0$) as $\dB$ grows to $\mathcal{A}\gg1$.

This method was motivated by the reduced equations of \alfreds, which 
describe the evolution of 1-D spherically polarised Alfv\'enic states in the  expanding solar-wind plasma. In this
case, the amplitudes of the fluctuating field and (radial) mean field evolve as $\propto a^{-1/2}$ and $\propto a^{-1}$, respectively,
where $a$ is the plasma's expansion factor. This leads to a similar situation where $\dB$ grows compared to $\mB$. Expansion 
also causes the gradient operators in \cref{eq: induction} to differ in the perpendicular 
and parallel directions and change with $a$, as well as rotating the
mean field if it has non-radial components. These effects necessitate extra terms in \cref{eq: induction,eq: phi constraint} (see 
\cref{sec: 1d} for details), but do not fundamentally modify the method. Such 
terms cause intriguingly  different nonlinear solutions to develop (see, e.g., figure 4 of \citealt{Squire2022})
pointing to interesting generalisations of our method, such as making the direction of $\mB$ change in time or modifying the gradient
operators to capture the physical effects of super-radial expansion and wind acceleration relevant to the inner heliosphere \citep{Tenerani2017}. 
The correspondence with \alfreds\ also shows that, while \cref{eq: induction,eq: phi constraint} are formulated on a purely 
mathematical basis, the process of $\dB$ growth does share important similarities 
with the real physical processes that generate large-amplitude spherically polarised states in the solar wind. 

\section{One dimension}\label{sec: 1d}

In one-dimensional solutions, $\dB$ varies only in the $\ph$ direction, which can be at an arbitrary angle to $\mB$. This 
case is significantly more straightforward than two or three-dimensional solutions and serves to 
illustrate some useful points. Denoting the co-ordinate in the $\ph$ direction as $\lambda$, such that $\nabla\phi = \ph d\phi/d\lambda = \ph\phi'$, where
$\ph $ is a unit vector, \cref{eq: induction,eq: phi constraint}  simplify to 
\begin{equation}
\frac{\partial\dB}{\partial t}= \dB - \frac{\partial}{\partial \lambda}\left[\phi' (\dB + \mB_{{\rm T}})\right],\quad |\dB +  \mB_{{\rm T}}|^{2} \phi'' = - \dB\cdot\mB.\label{eq: 1d version}
\end{equation}
Here $\mB_{{\rm T}} = \mB - \ph (\ph\cdot\mB)$ is the part of $\mB$ that is transverse to the mean field. We see a clear correspondence
with equations (59) and (61) of \alfreds\ (with $\mB_{{\rm T}}$ denoted $\bm{v}_{\rm AT}$), which 
can actually be equivalently derived by asserting that $\partial/\partial t\,\delta (B^{2})=0$ (i.e., in the same way as \cref{eq: induction,eq: phi constraint}), as
opposed to the 
asymptotic expansion in slow expansion rate used therein. The 
extra terms in \alfreds\ result from expansion induced rotation of $\ph$ and $\mB$, as mentioned above.

%%%%%%%%%%%%%%%%%%%%%%%%%%%%%%%%%%
\begin{figure}
\centering
\includegraphics[width=0.99\columnwidth]{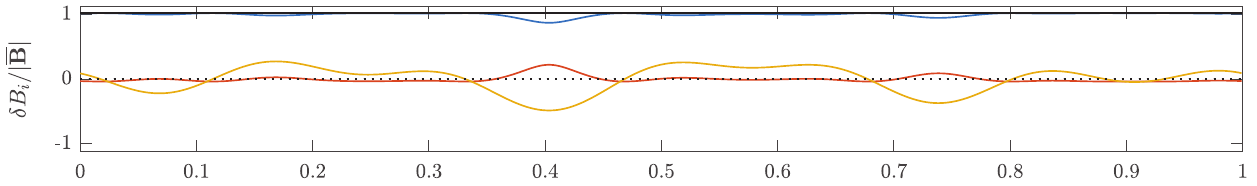}\vspace{-0.35cm}\\
\includegraphics[width=0.99\columnwidth]{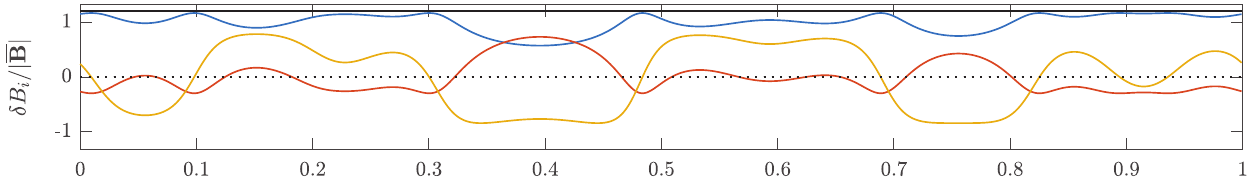}\vspace{-0.35cm}\\
\includegraphics[width=0.99\columnwidth]{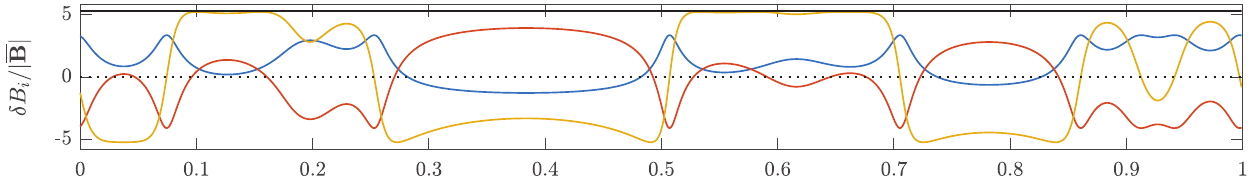}\vspace{-0.35cm}\\
\includegraphics[width=0.99\columnwidth]{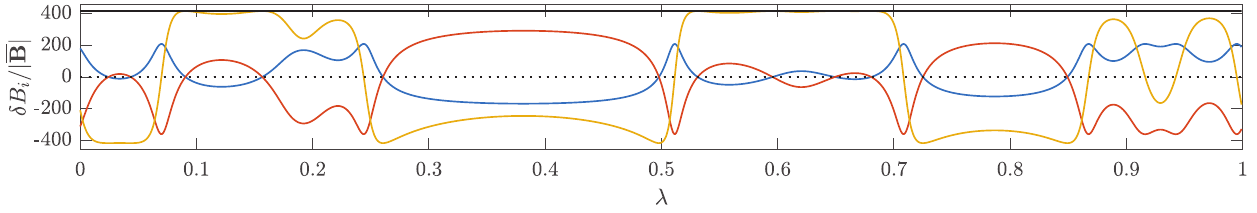}
\caption{One-dimensional spherically polarised solutions from \cref{eq: 1d version}, starting  from a solution of \cref{eq: bm solution} with $\mathcal{A}\approx 0.2$ constructed from
a random collection of the first 6 Fourier modes. The wavevector $\ph$ is at an angle of $30^{\circ}$ to $\mB$, which lies in the $\hat{x}$ direction. 
Colors show $B_{x}$ (blue), $B_{y}$ (red), $B_{z}$ (yellow), and $B$ (black). From top to bottom, the panels show $\mathcal{A}\approx 0.2$ (the initial conditions),  $\mathcal{A}\approx 0.65$, $\mathcal{A}\approx 5$,  and  $\mathcal{A}\approx 400$, illustrating how the solution approaches the zero-mean-field limit with $\dB\gg\mB$.}
\label{fig: 1d}
\end{figure}
%%%%%%%%%%%%%%%%%%%%%%%%%%%%%%%%%%

Solving \cref{eq: 1d version} first requires an initial condition with constant $B^{2}$. This is easily constructed by arbitrarily specifying
the component of $\dB$ in the $\hat{\bm{n}} \equiv \ph\times \mB/|\ph\times \mB|$ direction---i.e., the $\dB$ direction for a linear Alfv\'en wave---then solving for the $\dB$ component in the $\hat{\bm{m}}\equiv \ph\times \hat{\bm{n}}/|\ph\times \hat{\bm{n}}|$ direction, noting also that $\ph\cdot\dB=0$ 
due to periodicity in $\lambda$. This gives 
\begin{equation}
\hat{\bm{m}}\cdot\dB = - \hat{\bm{m}}\cdot\mB + \sqrt{B^{2} - (\ph\cdot\mB)^{2} - (\hat{\bm{n}}\cdot\dB)^{2}},\label{eq: bm solution}
\end{equation}
with  $\hat{\bm{n}}\cdot\dB$ an arbitrary function of $\lambda$. This must be combined with the condition $\overline{\hat{\bm{m}}\cdot\dB}=0$, which determines $B^{2}$. 
The solution \eqref{eq: bm solution}  illustrates two difficulties that also manifest in higher dimensions: \revchng{first, 
$B^{2}$ must be computed self-consistently for a particular form of $\dB$;
second, there may not exist real 
solutions for arbitrary choices of $ \hat{\bm{n}}\cdot\dB$ once $|\dB|$ approaches $|\mB|$.
If one is not careful, either of these difficulties will 
almost inevitably cause discontinuities  in $\dB$ as its solution is constructed.}
But, we reiterate that this does not
signify that spherically polarised solutions do not exist, or are non-smooth; 
rather, they require constraining  both components of $\dB$ simultaneously, as opposed to specifying one component and solving for the other. As a concrete example, \citet{Barnes1974} show that for $\hat{\bm{n}}\cdot\dB = \mathcal{A}_{n}\sin(k\lambda)$ solutions exist only for $\mathcal{A}_{n}<\mathcal{A}_{n,{\rm crit}}=(\pi/2)\sin\vartheta$, where $\vartheta = \cos^{-1}(\ph\cdot\mB/|\mB|)$. However, as we show below,
if one starts with such a solution and evolves it according to \cref{eq: 1d version}, there is no singular or otherwise interesting behaviour 
as $\mathcal{A}$ passes through $\mathcal{A}_{n,{\rm crit}}$; $\hat{\bm{n}}\cdot\dB$ simply changes shape to avoid the amplitude limit. 
This example illustrates why ``growing'' the $\dB$ solution from small amplitudes is necessary---without this, one is limited by the inability 
to choose an appropriate free function that will enable the constant-$B$ constraint to be satisfied.

\revchng{With a small-amplitude $\dB$ specified,} \cref{eq: 1d version} is  easily solved by standard numerical methods. Here we use sixth-order finite differences for consistency with the 2-D and 3-D calculations, but a Fourier pseudospectral method is equally suitable in 1-D (\alfreds; \citealp{Squire2022}). An example is shown in \cref{fig: 1d} at several  times as $\dB$ grows in 
amplitude. By the final snapshot, with $\mathcal{A}\approx 400$, the solution is approaching the zero-mean-field limit, which would give a purely stationary 
Alfv\'enic solution, since the propagation velocity $\va$ becomes much smaller than the flows $\delta \bm{u}$ in the nonlinear solution. Note that 
once $\dB\gg\mB$, $\phi$ becomes small and the shape of $\dB$ remains nearly constant with growing amplitude. 
%To our knowledge the properties of such states have not been studied. 

\section{Two dimensions}\label{sec: 2d}

To explore general 2-D solutions, we stipulate that $\dB$ is a function only of $\ell \equiv \hat{q}_{x} x  + \hat{q}_{y}y$ and $z$, taking  $\mB = \mBm\hat{\bm{x}}$. This geometry is the obvious generalization of the 1-D wave described above, with $\dB$ varying only in the plane angled at $\theta_{\rm 2D} \equiv \tan^{-1}(\hat{q}_{y}/\hat{q}_{x})$ to the mean field. 
As in 1-D, the 2-D calculation proceeds in two steps by first constructing a low-amplitude near-linear solution, then growing this using 
\cref{eq: induction,eq: phi constraint}. Both steps are significantly more complex than in 1-D.

\vspace{0.1cm}\noindent\textbf{Low-amplitude solution}~
First, we note that the clear generalization of the $\hat{\bm{n}}$-directed 1-D linear Alfv\'enic field to two (or three) dimensions
is the field $\dB = \nabla\times (A_{x}\hat{\bm{x}})$, with  $A_{x}$ chosen arbitrarily. The goal, then, is
to add  additional field components to enforce constant $B^{2}$, which is best done using the vector potential ($A_{y}$ and/or $A_{z}$) so as to maintain $\nabla\cdot\dB=0$ (the vector potential is not needed in 1-D because enforcing $\ph\cdot\dB=0$ ensures $\nabla\cdot\dB=0$).
The equation for $B^{2}$ becomes
\begin{equation}
B^{2} = {\rm const.} =  |\mBm-\partial_{z}A_{y} + \partial_{y} A_{z}|^{2} + |\partial_{z}A_{x} - \partial_{x}A_{z}|^{2} + |\partial_{x}A_{y} - \partial_{y}A_{x}|^{2},\label{eq: B2 vector pot}
\end{equation}
where $\partial_{x}=\hat{q}_{x}\partial_{\ell}$ and $\partial_{y}=\hat{q}_{y}\partial_{\ell}$. If one of $A_{y}$ or $A_{z}$ is fixed,
\cref{eq: B2 vector pot} is a first-order nonlinear partial differential equation (PDE) in $A_{z}$ or $A_{y}$, which can in principle be solved using 
the method of characteristics. However, such a method, which is effectively that used by  \citet{Valentini2019}, is problematic on 
a periodic domain because the value of the constant $B^{2}$ is not a-priori known, but itself depends on the solution ($A_{z}$ or $A_{y}$). An incorrect 
choice of $B^{2}$ manifests as an additional component of the mean field (as occurs for $\hat{\bm{m}}\cdot\dB$ in \cref{eq: bm solution}),
which will require $\bm{A}$ to contain linear gradients, and thus lead to spurious discontinuities on a periodic domain.
To overcome this, we instead stipulate that $\nabla B^{2}=0$ in \cref{eq: B2 vector pot}, which removes the issue of $B^{2}$ being undetermined at the cost of increasing
the order of the PDE. We also use the ``mean-field Coulomb gauge'' $A_{y} = -\partial_{z}\alpha$, $A_{z}=\partial_{y}\alpha$ so that \cref{eq: B2 vector pot} involves just one free function without causing an artificial difference between the $\ell$ and $z$ directions. We then solve the resulting nonlinear PDE for $\alpha(\ell,z)$ in Fourier space using MATLAB's \texttt{fsolve}
function with the trust-region dogleg method.  The two equations of $\nabla B^{2}=0$ are easily combined into one by minimizing only 
nonzero Fourier modes with \texttt{fsolve}, and the low-order approximate solution $(\partial_{y}^{2}+\partial_{z}^{2})\alpha = -(|\partial_{y}A_{x}|^{2} + |\partial_{z}A_{x}|^{2})/(2B_{0})$ provides a good initial guess for the optimization. So long as $A_{x}$ is chosen to be relatively smooth (e.g., 
the first one or two Fourier modes in each direction), the procedure is rapid and straightforward because only 
a small number of Fourier modes are needed to represent $\alpha$. We have found no evidence for the development of discontinuities in such solutions,
and a solution for $\alpha$ can be found for most choices of smooth, low-amplitude $A_{x}$ (at least in two dimensions; see below).

%%%%%%%%%%%%%%%%%%%%%%%%%%%%%%%%%%
\begin{figure}
\centering
\includegraphics[width=1.0\columnwidth]{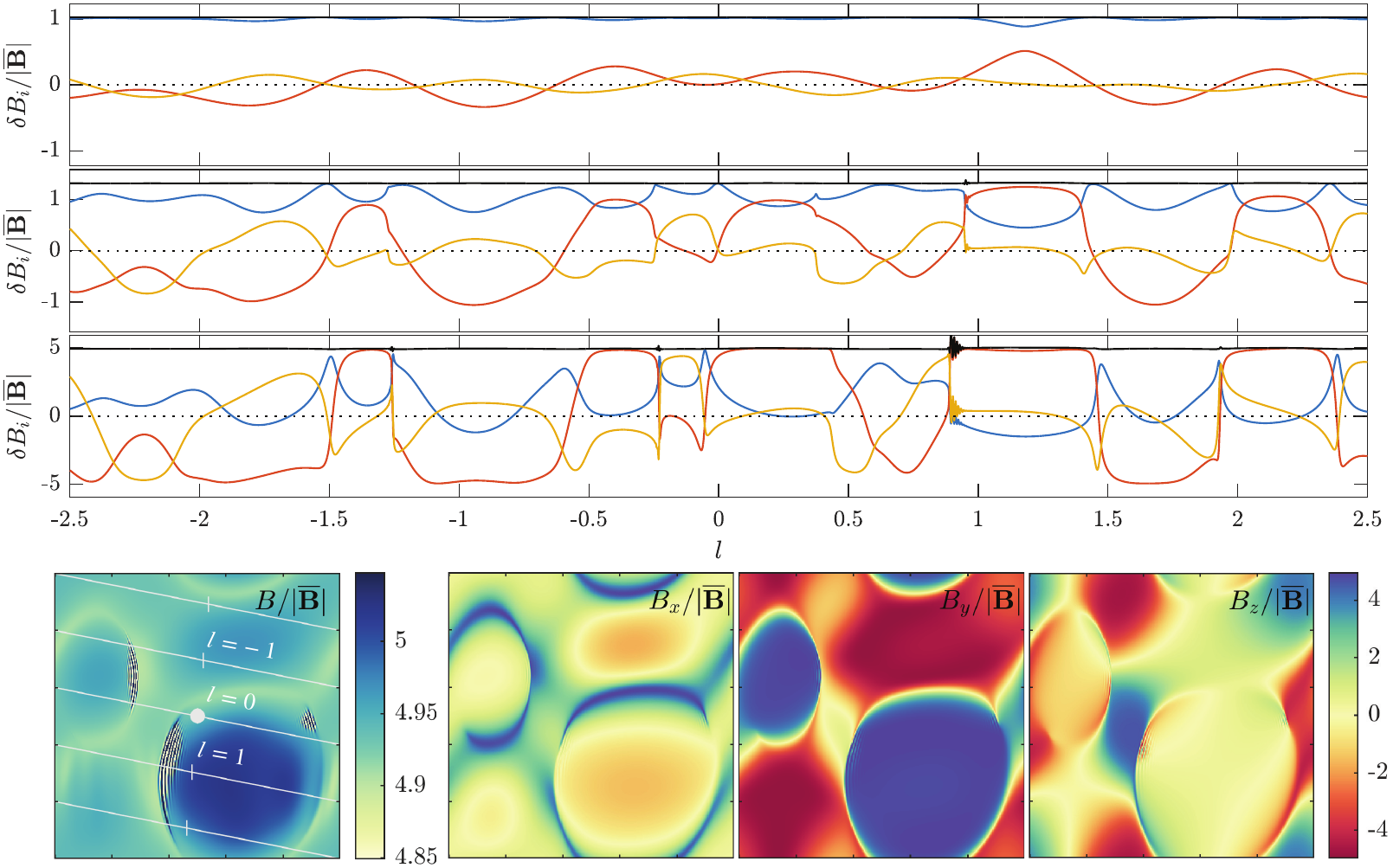}
\caption{Two-dimensional spherically polarised solution on an $\ell,z$ plane angled at $\theta_{\rm 2D}=30^{\circ}$ from 
the $\hat{\bm{x}}$ (mean-field) direction. The top three panels show periodic traces of $B$ (black), $B_{x}$ (blue), $B_{y}$ (red), and $B_{z}$ (yellow)
along the white line plotted on the bottom-left panel (this lies at angle $\alpha\approx 11.3$ degrees from  $\hat{\bm{\ell}}$; \revchng{integer $l$ values are marked to illustrate the correspondence between the trace and the 2-D solution}).
The initial condition is constructed from a random superposition of modes in $A_{x}$, scaled to give amplitude $\mathcal{A}\approx 0.2$ (top panel). It then 
grows in time according to \cref{eq: induction,eq: phi constraint}. The bottom panels show the 2-D structure of the components of 
$\bm{B}$ at the time corresponding the bottom trace, when $\mathcal{A}\approx 5$. 
At least to the precision of the $384^{2}$ resolution  used here, discontinuities develop in the field structure, unlike the 1-D solutions (the
most prominent is near near $l=1$ on the trace plots).
Aside from numerical ringing caused by the development of these discontinuities, however, 
$B$ remains very constant throughout the domain (the colorbar of $B$ on the bottom left is scaled to $\pm 2\%$) }
\label{fig: 2d}
\end{figure}
%%%%%%%%%%%%%%%%%%%%%%%%%%%%%%%%%%

\vspace{0.1cm}\noindent\textbf{Large-amplitude solutions}~
Starting from a constant-$B$ initial condition, \cref{eq: induction,eq: phi constraint} can be solved using standard methods. For simplicity,
we use an Euler timestepper with the timestep chosen based on the maximum of $\phi$ over the domain.
The only complication arises from the non-homogenous derivative operator in the Poisson equation \eqref{eq: phi constraint}, which
must be recomputed and solved at every timestep as $\dB$, and thus $\nabla_{\perp}$, change. We choose to use a periodic sixth-order finite-difference 
representation\footnote{A range of finite-difference orders were tested. Lower-order operators cause a clear decrease in solution quality, but there was little gained
for orders above $6$. },
forming $\nabla_{\perp}^{2}$ as a sparse matrix through left-multiplication of the relevant gradient operators by $B_{i}$.
A difficulty in the interpretation and solution of \cref{eq: phi constraint} is that the $\nabla_{\perp}^{2}$ matrix has a rather
high dimensional nullspace, which means that,  depending on the source $-\dB\cdot\mB$, \cref{eq: phi constraint} may not have a 
solution\footnote{To understand why this is the case, it is helpful to consider a simple example such as $\bm{B}=\hat{\bm{z}}$, which gives $\nabla_{\perp}^{2}=\partial_{\ell}^{2}$. Taking arbitrary $f$ in $\nabla_{\perp}^{2}\phi = f$, it is clear that any $\ell$-independent $f$ that is periodic in $z$ cannot be captured by $ \nabla_{\perp}^{2}\phi$ on a periodic domain. Thus, the solution to \cref{eq: phi constraint} cannot capture any part
of $\dB\cdot\mB$  that varies along $\bm{B}$ but is constant perpendicular to it (though, exactly what this statement means for 
a complex 2-D or 3-D $\dB$ is not obvious). Note that this difficulty is absent in 1-D because all qualities vary only in $\lambda$.}.
We circumvent the issue by interpreting \cref{eq: phi constraint} in the least-squares sense, thus finding the best approximation to 
the flow $\tilde{\bm{u}}=\nabla\phi$ that maintains constant $B$. We use the least-squares conjugate-gradient
method \citep{Paige1982} with an incomplete LU preconditioner. As we show below, the solution is generally very 
good, but the method certainly does not guarantee constant $B$, at least at finite resolution, unlike in 1-D.  Presumably, if   $-\dB\cdot\mB$ has a large
component in the nullspace of $\nabla_{\perp}^{2}$, this implies that $\dB$ has developed structures that cannot continue to grow in 
amplitude while maintaining constant $B$. Understanding the conditions under which this occurs requires further study. 

%%%%%%%%%%%%%%%%%%%%%%%%%%%%%%%%%%
\begin{figure}
\centering
\includegraphics[width=1.0\columnwidth]{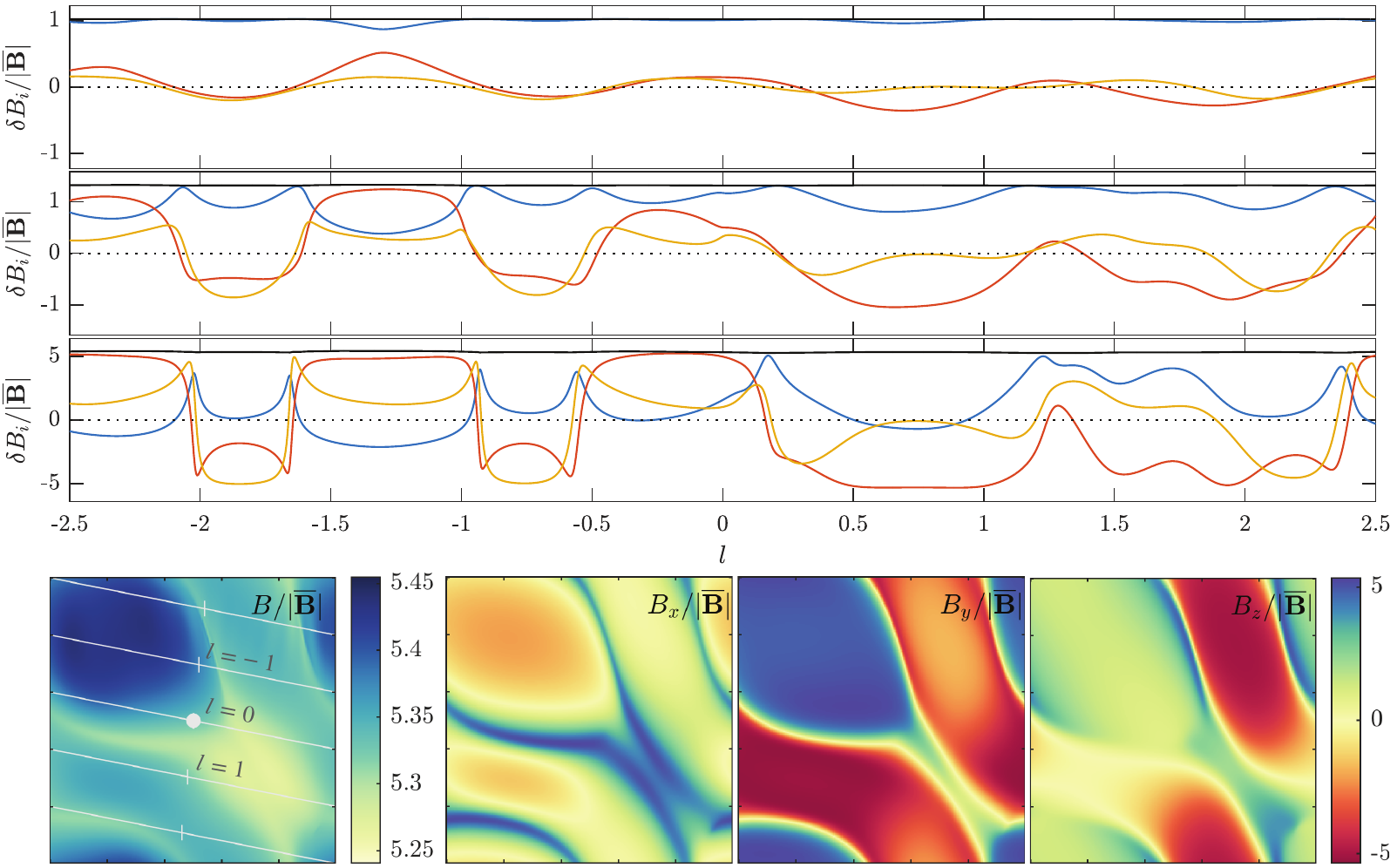}
\caption{Same as \cref{fig: 2d} but starting from a different initial random collection of Fourier modes in the 
low-amplitude $A_{x}$ initial conditions. We show only the solution with $\mathcal{A}\approx 5$. In this case, discontinuous structures do not develop and the solution is well resolved 
at this resolution of $256^{2}$ (which is lower than \cref{fig: 2d})  }
\label{fig: 2d 2}
\end{figure}
%%%%%%%%%%%%%%%%%%%%%%%%%%%%%%%%%%

An example solution is shown in \cref{fig: 2d}. We use a periodic domain of size $L_{\ell}=1$, $L_{z}=1$, with $384$ grid points in 
each direction and $\theta_{\rm 2D}=30^{\circ}$. 
The initial conditions in $A_{x}$ are constructed from 
a random combination of $|k_{\ell}|\leq 2\pi/L_{\ell}$ and $|k_{z}|\leq 2\pi/L_{z}$ modes, scaled to give $\mathcal{A}\approx0.2$.
In the top three panels, we show the field structure by taking an angled trace through the domain, 
accounting for the periodicity in order to show most of the solution \revchng{(but note that some larger structures appear twice; the
bottom-left panel illustrates the correspondence to the 2-D domain)}. The smooth, small-$\delta B_{x}$ initial conditions are shown in the
top panel of \cref{fig: 2d} and have a very small variation in $B$, with $({\overline{B^{2}}})^{1/2}/\overline{B}\approx 10^{-5}$. 
As $\dB$ grows, it develops quite sharp gradients by modest amplitudes ($\mathcal{A}\gtrsim0.7$; see second panel), which are 
are numerically unresolved at this resolution. The large-amplitude $\mathcal{A}\approx 5$
solution is shown both in bottom trace and in the image plots in the bottom row, illustrating the discontinuities  that have developed at several locations in the domain. 
These sharp gradients represent a distinct difference compared to 1-D solutions, which only ever steepen modestly from the 
initial waveform as they grow to large 
amplitudes (see \cref{fig: 1d}; this is discussed in more detail below).
%Interestingly, a different initial $A_{x}$, shown in \cref{fig: 2d 2}, does not develop gradients and is well resolved at $256^{2}$
The method clearly 
maintains extremely constant $B$, aside from numerical ringing near some of the sharp gradient discontinuities that form (excluding 
these regions $({\overline{B^{2}}})^{1/2}/\overline{B}\approx 0.8\%$).

We have also explored solutions on planes at other angles $\theta_{\rm 2D}$ (not shown) and with differing initial conditions. 
As we demonstrate in  \cref{fig: 2d 2}, which shows another example large-amplitude solution,
the propensity to form
sharp gradients  depends on the structure of the initial low-amplitude $A_{x}$ (see below for further discussion). Unlike \cref{fig: 2d}, the solution in \cref{fig: 2d 2} is well resolved and relatively smooth, constituting a practical demonstration that 2-D smooth, 
spherically polarised solutions exist (which, as far as we are aware, was not 
previously known). 
We have not found any simple heuristic to determine how the 
sharpness of the final solution relates to the initial low-amplitude one. 
In general, 
solutions with larger $\theta_{\rm 2D}$---\emph{viz.,} those which are more elongated along the magnetic field--- develop somewhat larger variation in $B$. This seems to be due to 
larger inaccuracies in the solutions of \cref{eq: phi constraint}, although whether this relates to discontinuities remains unclear.

\section{Three dimensions}\label{sec: 3d}

\begin{figure}
\centering
\includegraphics[width=1.0\columnwidth]{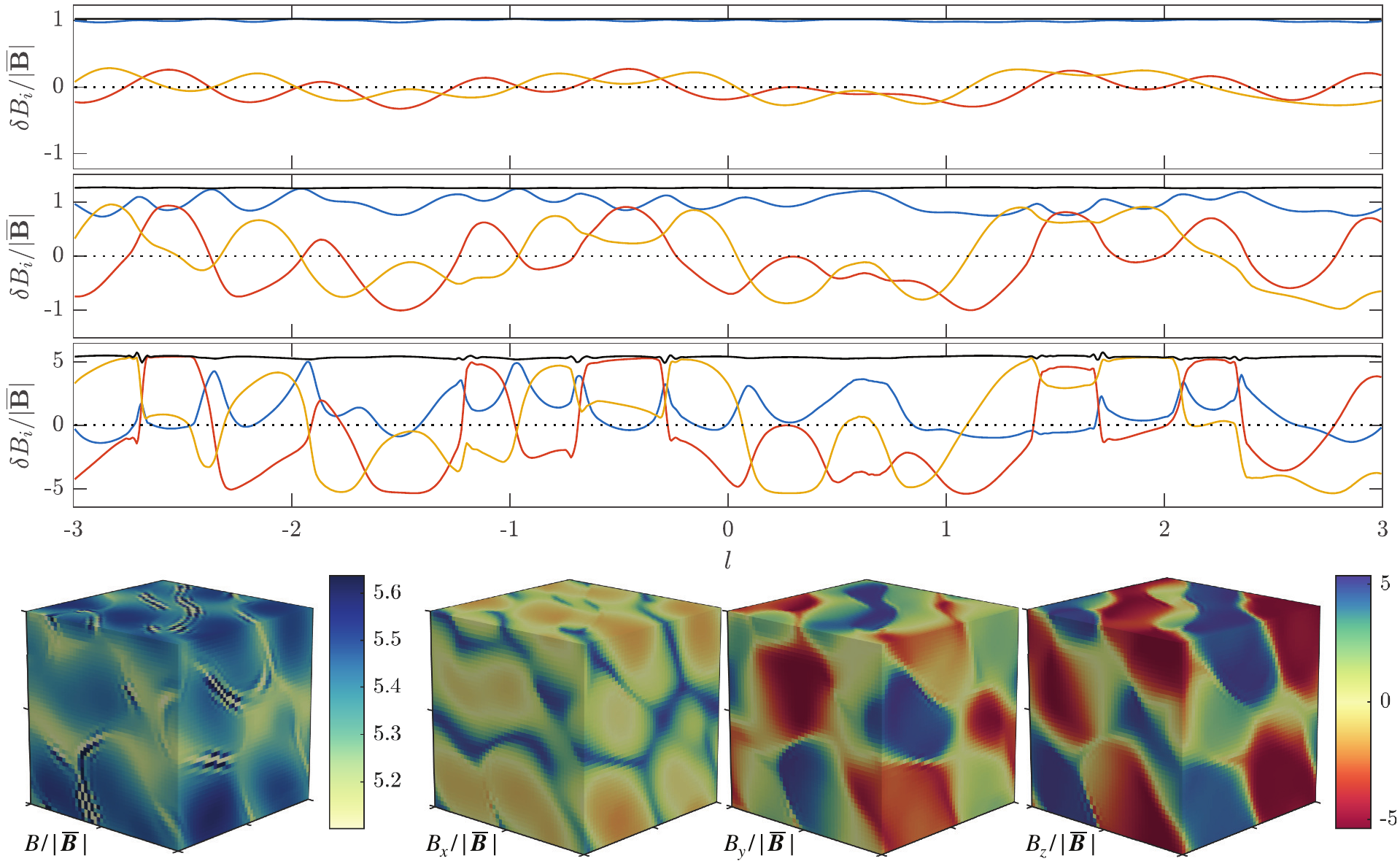}
\caption{Three-dimensional spherically polarised solution in a cubic box with a resolution of $48^{3}$. As in \cref{fig: 2d}, the top three panels show periodic traces of $B$ (black), $B_{x}$ (blue), $B_{y}$ (red), and $B_{z}$ (yellow) along a line in the direction $(\cos \theta_{\rm 3D}, \sin \theta_{\rm 3D}\,\cos\varphi_{\rm 3D},\sin \theta_{\rm 3D}\,\sin\varphi_{\rm 3D})$, with $\theta_{\rm 3D}\approx 30^{\circ}$ and $\varphi_{\rm 3D}\approx 11.3^{\circ}$, with $l=0$ at the center of the 
box (the units are scaled to the size of the box).
The initial condition is constructed from random modes in $A_{x}$ with an amplitude such that $\mathcal{A}\approx 0.2$ (top panel), then 
growing in time according to \cref{eq: induction,eq: phi constraint}. The bottom panels show the 3-D structure of the components of 
$\bm{B}$ at the time corresponding the bottom trace, when $\mathcal{A}\approx 5$. 
}
\label{fig: 3d}
\end{figure}

Three-dimensional solutions are constructed through a method that is almost identical to the 2-D case. 
The only additional complication is that the method to construct the low-amplitude 
solution using the mean-field Coulomb gauge ($A_{y} = -\partial_{z}\alpha$, $A_{z}=\partial_{y}\alpha$) cannot
be used to construct a field that varies in $x$ but not in $y$ or $z$ (i.e., a field with power in  $k_{\perp} = \sqrt{k_{y}^{2}+k_{z}^{2}}=0$ modes). 
Although a different method of constructing a low-amplitude solution may alleviate this, we opt instead 
to adjust the chosen $A_{x}$ so that its associated $B^{2} = |\partial_{y}A_{x}|^{2} + |\partial_{z}A_{x}|^{2}$ has very little power in $k_{\perp}=0$ modes.
This is easily done using MATLAB's \texttt{lsqnonlin} function after constructing $A_{x}$ from a collection of random Fourier modes. 
The method seems to work well; for example, it generates  smooth initial conditions with $\mathcal{A}\approx 0.2$ and $({\overline{B^{2}}})^{1/2}/\overline{B}\approx 3\times 10^{-5}$ (see \cref{fig: 3d}). In solving \cref{eq: induction,eq: phi constraint}, the only extra challenge
compared to 2-D is the computational expense, and we are limited to constructing solutions
with resolutions ${\lesssim}48^{3}$ due to our non-parallelised implementation using MATLAB. However, the only significant computational difficulty---
inverting $\nabla_{\perp}^{2}$ in \cref{eq: phi constraint}---involves standard iterative matrix-solve methods, which  have robust and efficient parallel 
implementations that could allow for much higher resolutions if desired. 
%%%%%%%%%%%%%%%%%%%%%%%%%%%%%%%%%%
%%%%%%%%%%%%%%%%%%%%%%%%%%%%%%%%%%

An example 3-D solution in a cubic box is shown in \cref{fig: 3d}. As in \cref{fig: 2d}, we show the solution structure using a trace along an angled, periodic line, at $\mathcal{A}\approx0.2$ (the initial conditions), $\mathcal{A}\approx 0.7$, and $\mathcal{A}\approx 5$. Like in 2-D, the solutions appear to become discontinuous, although it is more difficult to diagnose in detail  because of the limited resolution. 
The solutions have somewhat larger variation in $B^{2}$ compared to \cref{fig: 2d}, with $({\overline{B^{2}}})^{1/2}/\overline{B}\approx 3\%$ by $\mathcal{A}\approx 5$, although
this is at least partially a result of the lower resolution.
We have also explored solutions in boxes that are elongated along $\mB$ with $L_{x}=4$, finding similar structures and properties, 
at least to the accuracy achievable here.
We also note that we have confirmed  that $\nabla\cdot\bm{B}$ remains zero within the tolerances of the finite difference representation (with  sixth-order finite differences at $48^{3}$ the root-mean-square of $\nabla\cdot\delta \bm{B}$ is ${\approx} 1.5\times 10^{-6}$).

\section{Discussion and conclusions}\label{sec: discussion}

% Key poitns
% Not guaranteed to form a constant-B solution in 2-D or 3-D. But, if it does, then clearly such solutions exist. 
% Interesting for two reasons. (i) to use as input for other calcs, interesting to explore different examples, e.g., spectra
% (ii) properties of the solutions. Clear that much sharper gradients form in high dimensions. Dependence on ICs suggests that would be resolved at higher res
% new para: observations & application to SBs

Above, we have presented a method to construct large-amplitude spherically polarised Alfv\'enic structures, which form a broad class of nonlinear solutions to the compressible MHD (or kinetic MHD) equations.
The method works by ``growing'' a low-amplitude, nearly linear ($\dB\ll \mB$) wave, thus allowing the exploration of 
nonlinear solutions in the large-amplitude $\dB\sim \mB$ regime applicable to the solar wind, or even  zero-mean-field limiting solutions with 
$\dB\gg \mB$.
We have presented some examples of such solutions in one, two, and three dimensions, albeit with somewhat limited resolution 
due to computational challenges. Others---including, in principle, solutions with turbulent-like spectra\footnote{Reliably doing this 
would require a more refined optimization method for generating the low-amplitude solution, as described in \cref{sec: 2d}; our current implementation can work reliably only with a relatively smooth choice for $A_{x}$.  }---could be generated using our method by choosing a different form for
the initial, small-amplitude wave. \revchng{These solutions exhibit some important  features observed in the solar wind, 
particularly the asymmetry (one-sidedness) of  $\delta B_{\|}=\dB\cdot\mB/|\mB|=\delta B_{x}$ fluctuations at modest $\mathcal{A}$, which is a 
consequence of maintaining constant $B$ through large changes in $\dB $ \citep{Gosling2009}. 
At  larger $\mathcal{A}$, all solutions cause magnetic-field reversals ($|\delta B_{\|}|>|\mB|$, a commonly used definition of switchbacks), 
although these effects would be stronger if we chose more oblique solutions (e.g., larger $\theta_{\rm 2D}$ or a more elongated box in 3-D; \citealp{Mallet2021}).}
However, it is also worth noting that despite its apparent success in nearly all 
cases we have explored, the method is not guaranteed to produce perfectly constant-$B$ structures because the Poisson-like 
equation \eqref{eq: phi constraint}, which determines how $\dB$ changes shape as it grows, lacks exact solutions in general, at least at finite resolution. These mathematical issues should be explored in more detail in future work. 
Other possible ideas for future studies include using a time-dependent mean field or gradient operators, which
would change the large-amplitude $\dB$ that results from chosen small-amplitude initial conditions (\alfreds; \citealp{Squire2022}), or the extension  to the relativistic regime \citep{Mallet2021a}, which may have interesting
applications to a number of high-energy processes such as disk coronae \citep{Chandran2018a} or pulsar magnetospheres \citep{Kumar2020,Zhang2020}.

One interesting use case for these solutions is as input 
for nonlinear MHD or kinetic simulations to study processes such as parametric decay (instability of Alfv\'en waves; \citealp{Sagdeev1969}) or large-amplitude reflection-driven turbulence \citep{Johnston2022}. 
\revchng{For example, a 3-D MHD simulation could be initialised with a 1-D, 2-D, or 3-D solution $\dB$ from our method by first extending it via symmetry to three dimensions (e.g., $\dB_{\rm 3D}(x,y,z) = \dB(\ph\cdot \bm{x})$ for a 1-D solution), 
then setting $\delta \bm{u} = \pm\delta \bm{B}/\sqrt{4\pi \rho}$.}  Study of parametric decay
  especially
has been somewhat limited by a lack of general,  large-amplitude \revchng{Alfv\'enic solutions other than circularly polarized waves} that can be evolved numerically, so 
our method may  enable important progress in this area  (the modest-amplitude case with $\dB\lesssim \mB$ has been studied in 1-D by \citealt{DelZanna2001,DelZanna2015}; \revchng{\citealt{Tenerani2020} considered a larger-amplitude 2-D solution}). A particularly interesting regime, which could in principle be studied for solutions of any dimensionality,
would be the zero-mean-field limit, where the nonlinear Alfv\'enic solutions become a stationary, non-propagating tangle of magnetic field lines and Alfv\'enic flows. The system is reminiscent of the tangled fields possible in magnetostatic force-free equilibria (those with $ \bm{u}=0$ and $\bm{B}\times (\nabla \times\bm{B})=0$; e.g., \citealt{Chandrasekhar1958,Marsh1996,Hosking2020}), but with a different class of equilibria that includes flows and has seen comparatively little study.

%%%%%%%%%%%%%%%%%%%%%%%%%%%%%%%%%%
\begin{figure}
\centering
\includegraphics[width=0.346\columnwidth]{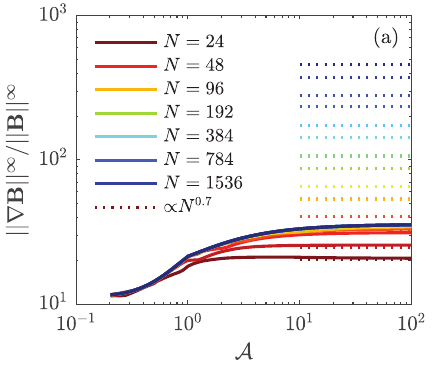}\includegraphics[width=0.327\columnwidth]{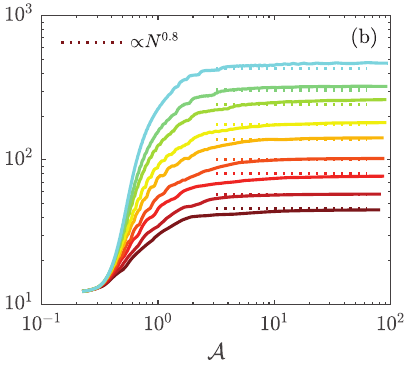}\includegraphics[width=0.327\columnwidth]{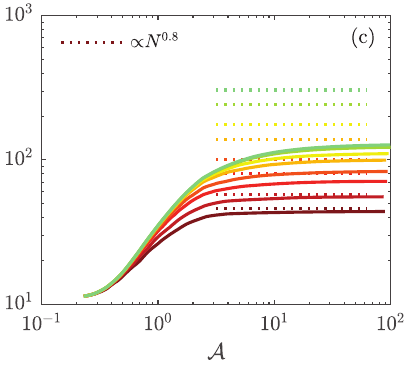}
\caption{Measurement of discontinuity formation in 1-D solutions (panel a) and 2-D solutions starting from two different initial conditions (panels b and c). We 
plot the normalised infinity norm of the gradient of the solutions,  $||\nabla \B||^{\infty}/||\B||^{\infty} = (1/3)\sum_{i}\sum_{j}\max(\nabla_{j} B_{i})/\max(B_{i})$ (with the $\nabla_{j}$ taken along either $\lambda$ or $\ell$ and $z$),  as a function
of $\mathcal{A}$ for a scan in resolution in each case (an $N\times N$ grid is used in 2-D; we list only every second $N$ in the legend for clarity).  In 1-D (left panel), we initialise with a linear 
combination of modes with $k\leq 4\pi/L$; in 2-D, we initialise with a linear combination of modes with $k_{\ell}\leq 2\pi/L_{\ell}$ and $k_{z}\leq 2\pi / L_{z}$
(the $N=384$ case of panel b is that shown in \cref{fig: 2d}). Dotted lines in each case show a scaling $N^{\chi}$,  with $\chi$ chosen to match the scaling of unconverged  solutions ($\chi\approx 0.7$ in 1-D and $\chi \approx 0.8$ in 2-D). Clearly, the 1-D solution converges at very low resolution ($N\approx 64$), showing that
\cref{eq: induction} does not lead to particularly small-scale features. In contrast, in 2-D, sharp field 
structures form much more readily: in the first example in panel (b), which is that from \cref{fig: 2d}, there is no convergence
even at the 
highest resolution 
that is feasible using our current computational implementation ($N=384$); but, the second example in panel (c), which is that from 
\cref{fig: 2d 2} and simply starts 
from a different random initial condition, achieves convergences around $N=128$.  }
\label{fig: discon}
\end{figure}
%%%%%%%%%%%%%%%%%%%%%%%%%%%%%%%%%%

A second reason for interest in these  solutions relates directly to their properties and structure. As far as we are aware, it was 
previously not known whether smooth, spherically polarised solutions with two- or three-dimensional structure even existed; our 
method has provided  near-perfect examples in 2-D, 
one of which is shown in  \cref{fig: 2d 2}. In 3-D, we are resolution limited due to our non-parallelised numerical implementation (the example in  \cref{fig: 3d} is not smooth at this resolution), but it seems unlikely that it would differ fundamentally 
from  2-D. Interestingly, in some other solutions, an example of which is shown in \cref{fig: 2d}, $\B$ develops extremely sharp gradients that remain discontinuous at our highest  resolution of $384^{2}$. 
We demonstrate this in \cref{fig: discon}, which quantifies the development of sharp structures by plotting how
the maximum gradient of  $\bm{B}$ evolves with solution amplitude, over a scan in resolution. For reference, we plot
with dotted lines the empirically determined power-law behaviour of an unconverged solution ($||\nabla \B||^{\infty}/||\B||^{\infty} \propto N^{\chi}$; the exponent $\chi$ differs between 1-D and 2-D). Convergence occurs at low resolution in 1-D solutions (\cref{fig: discon}a), which only ever develop modestly sharper structures compared to the  initial wave, and are converged by $N\approx 48$ for this
example ($b_{n}$ initialised with $k\leq 4\pi/L$ modes). In 2-D, both solutions
become significantly sharper, but panel (b) (that from \cref{fig: 2d}) shows no sign of convergence at all by $384^{2}$, while panel (c) 
(the case in \cref{fig: 2d 2}) converges at around $128^{2}$.
Given there does not seem to be any simple distinguishing feature(s) that determine whether discontinuities develop\footnote{The solution in \cref{fig: 2d 2} could be considered ``more 1-D'' than that in \cref{fig: 2d}, in that in varies predominantly along a diagonal line
across the domain. However, some other cases explored do not share this property and  still do not develop sharp gradients, while 
other ``nearly 1-D'' solutions do seem to become extremely sharp. }, the 
most likely scenario seems to be that continuous solutions do generally exist, but they require extremely sharp gradients in some cases. If true, this suggests 1-D waves are simply a particular special case of  more general 2-D or 3-D fields that happens to allow  large-amplitude solutions with an especially smooth structure.

The conclusions of the previous paragraph and \cref{fig: discon} may have interesting consequences for switchback formation in 
the solar wind. Specifically,
our method based on \cref{eq: induction,eq: phi constraint} shares clear  similarities with the \emph{physical} processes that occur in the 
solar wind, where Alfv\'enic structures grow  in normalised amplitude due to plasma expansion. Although the direct effect of expansion is not included in \cref{eq: induction,eq: phi constraint} (this would cause time dependence of the gradient operators), the general idea---whereby, in order to 
maintain constant $B^{2}$ as it grows, $\dB$ changes shape  by means of a compressive flow---is very similar to physical expansion, and would presumably produce similar $\dB$ evolution\footnote{This ignores
the effect of wave reflection in creating backwards propagating waves that seed turbulence, which  
creates smaller-scale structures in a very different way. In ignoring such effects, 
we assume the system
remains strongly dominated by outwards-propagating fluctuations and thus can remain nearly  Alfv\'enic.}.
In support of this idea, in 1-D with expansion effects added, \cref{eq: induction,eq: phi constraint} become exactly the system of \alfreds, which 
was derived asymptotically from the expanding MHD equations and produces similar large-amplitude structures to \cref{eq: induction,eq: phi constraint}.
If this correspondence holds in 3-D, our results imply that small-amplitude Alfv\'enic perturbations released from the low corona would often  develop  
very sharp gradients or discontinuities in the process of growing, unless they are in some sense one dimensional (unlikely if
they are created by turbulence in the chromosphere; \citealp{vanBallegooijen2011}). 
%Such a process would occur alongside any development of turbulence from wave reflection, so long as the system remained strongly dominated by outwards-propagating fluctuations. 
This is quite promising for the \emph{in-situ} Alfv\'enic scenario of 
switchback formation \citep{Squire2020,Shoda2021}, in which switchbacks are 
simply Alfv\'enic fluctuations that have grown to large amplitudes by expansion. 
While \alfreds\  suggested, based on 1-D solutions,  that the model may struggle to explain
the extremely sharp switchback structures seen in some observations \citep{Kasper2019,Bale2019,AkhavanTafti2021},
the answer may simply be that near-discontinuities naturally develop when starting from low-amplitude
fluctuations with more general (non 1-D) structure. \revchng{Whether this idea has direct observable 
consequences---for example, in the relative populations of rotational versus tangential discontinuities \citep[e.g.,][]{Neugebauer1984}---requires better understanding of why and how sharp gradients develop, so is left to future work.     }

\begin{acknowledgements}The authors acknowledge useful discussion 
with J.~Burby, R.~Meyrand, B.~Chandran,  Z.~Johnston, and J.~N\"attil\"a in relation to this work. Support for  J.S. was
provided by Rutherford Discovery Fellowship RDF-U001804, which is managed through the Royal Society Te Ap\=arangi. 
A.M. acknowledges the support of  NASA through  grant 80NSSC21K0462. 
\end{acknowledgements}

Declaration of Interests: The authors report no conflict of interest.

%\bibliographystyle{jpp}
%\bibliography{fullbib_formatted}

\end{document}